\documentclass[letterpaper]{article}

\usepackage[T1]{fontenc}
\usepackage{mathpazo}
\usepackage{amsmath}
\usepackage{amssymb}
\usepackage{hyperref}
\usepackage{graphicx}
\usepackage{color}
\usepackage{enumitem}
\usepackage{booktabs}
\usepackage{geometry}
\usepackage{titlesec}
\usepackage{marginnote}

\hypersetup{
    colorlinks=true,
    linkcolor=blue,
    urlcolor=blue,
    citecolor=blue
}

\titleformat{\section}{\large\bfseries}{\thesection}{1em}{}
\titleformat{\subsection}{\normalsize\bfseries}{\thesubsection}{1em}{}

\title{\textbf{Why iCloud Fails:}\\[0.3em]
{\large The Category Mistake of Cloud Synchronization,\\
and Why OAE Transactional Semantics Can Resolve It}}

\author{Paul Borrill\\
\small DAEDAELUS\\
\small\texttt{paul@daedaelus.com}\\
\small ORCID: 0000-0002-7493-5189}

\date{Version 0.7 \quad 02026-MAR-07}

\begin{document}

\maketitle

\begin{abstract}
iCloud Drive presents a filesystem interface but implements cloud synchronization semantics
that diverge from POSIX in fundamental ways. This divergence is not an implementation bug;
it is a \emph{Category Mistake}---the same one that pervades distributed computing wherever
Forward-In-Time-Only (FITO) assumptions are embedded into protocol design. Parker et al.\
showed in 1983 that network partitioning destroys mutual consistency; iCloud adds a user
interface that conceals this impossibility behind a facade of seamlessness. This document
presents a unified analysis of why iCloud fails when composed with Time Machine, git,
automated toolchains, and general-purpose developer workflows, supported by
direct evidence including documented corruption events and a case study involving 366\,GB
of divergent state accumulated through normal use. We show that the failures arise from
five interlocking incompatibilities rooted in a single structural error: the projection of
a distributed causal graph onto a linear temporal chain. We then show how the same Category
Mistake, when it occurs in network fabrics as link flapping, destroys topology knowledge
through epistemic collapse. Finally, we argue that Open Atomic Ethernet (OAE) transactional
semantics---bilateral, reversible, and conservation-preserving---provide the structural
foundation for resolving these failures, not by defeating physics, but by aligning protocol
behavior with physical reality.
\end{abstract}

\section{Introduction: iCloud Is Not a Filesystem}

Users are encouraged to believe that iCloud Drive is a transparent extension of the local
filesystem. Apple's marketing promises seamless synchronization: files appear on every
device, changes propagate automatically, and the experience is indistinguishable from
local storage.

This promise routinely fails to deliver. Not because Apple's engineers are incompetent,
but because the promise itself rests on a Category Mistake.

A filesystem provides:
\begin{itemize}[nosep]
  \item stable paths that do not change based on operational state,
  \item atomic operations (rename, create, link) with POSIX semantics,
  \item advisory locking that coordinates concurrent access,
  \item consistent directory listings that reflect recent writes,
  \item the guarantee that listed files are present and readable.
\end{itemize}

iCloud Drive can violate each of these properties under documented conditions:
\begin{itemize}[nosep]
  \item File paths change based on sync state (legacy \texttt{.icloud} stubs; modern dataless files),
  \item Operations are not atomic across the sync boundary,
  \item \texttt{NSFileCoordinator} replaces POSIX locking with optional, cooperative IPC,
  \item Directory listings include dataless files with no data extents,
  \item Files may be listed but require network fetches before content is available.
\end{itemize}

iCloud Drive is a \emph{distributed negotiation system} that presents a filesystem facade.
Treating this facade as authoritative is the Category Mistake that produces the failures
users experience daily.

The consequences extend beyond inconvenience. As Dominik Mayer documented, iCloud Drive
silently deletes user content when version conflicts arise
\cite{mayer2023icloud-deletes}. Mayer identifies the core of the problem as a
\emph{product decision}: ``the core issue is the product decision to not be transparent
about version conflicts. There is no central interface to manage them, nor are they
somehow represented on the hard drive so you could deal with them whenever you are
looking at your files.'' The system does not merely fail to synchronize correctly; it
actively conceals the fact that it has failed.

This is not a new problem. Parker et al.\ established in 1983 that network partitioning
can ``completely destroy mutual consistency'' in replicated filesystems
\cite{parker1983detection}. The theoretical impossibility was clear forty years ago.
What iCloud adds is a user interface that hides the impossibility behind a facade of
seamlessness, converting a known hard problem into silent data destruction.

\section{The Structural Thesis}

The failures of iCloud, when composed with Time Machine, git, automated toolchains, or any
program that assumes POSIX semantics, are not random bugs. They are \emph{projection artifacts}
caused by a single structural error:

\begin{quote}
\emph{Projecting a distributed causal graph onto a linear temporal chain destroys
essential structure. Information is lost. The losses manifest as corruption, conflicts,
stalls, and silent data destruction.}
\end{quote}

This is an instance of Forward-In-Time-Only (FITO) thinking \cite{borrill2025wdcgw}.
FITO replaces partial order with total order and treats the result as authoritative
history. Cloud sync systems operate in causal order. Snapshot systems, timestamp-based
conflict resolution, and lock-file protocols operate in wall-clock order. These are not
equivalent.

Parker et al.\ showed in 1983 that mutual consistency cannot be maintained under
network partitioning \cite{parker1983detection}. Every cloud sync system---iCloud,
Google Drive, Dropbox, OneDrive---operates in exactly this regime: intermittently
connected devices that partition whenever a laptop lid closes or a phone enters a
tunnel. The failures documented below are not edge cases. They are the predicted
consequences of a problem the distributed systems community identified and
characterized four decades ago.

\section{The Architecture of iCloud Drive}
\label{sec:architecture}

To understand why iCloud fails, we must understand what it actually is.

\subsection{The Coordination Layer}

iCloud Drive does not use POSIX advisory locks. It employs \texttt{NSFileCoordinator},
a higher-level coordination mechanism that mediates access between applications and the
\texttt{bird} sync daemon via inter-process communication \cite{apple-nsfilecoordinator}.
This coordination is \emph{cooperative and optional}: any process can bypass it by
operating directly on files. The sync daemon itself has exclusive access during
upload/download operations, producing the notorious ``Operation not permitted'' errors
when tools like \texttt{rm} attempt to modify syncing files.

\subsection{Temporal Semantics: Last-Writer-Wins}

iCloud resolves conflicts using modification timestamps---last-writer-wins. It does
\emph{not} use vector clocks or logical timestamps \cite{apple-tn2336}. This means:
\begin{itemize}[nosep]
  \item clock skew across devices produces false ordering,
  \item ``later'' is treated as ``more correct,''
  \item a single linear history is assumed where none exists.
\end{itemize}

None of these assumptions hold reliably in a distributed, intermittently connected
environment.

\subsection{CloudKit and FoundationDB}

The server side runs on CloudKit, which is built on FoundationDB's Record Layer---an
ACID-compliant, multi-tenant distributed database \cite{foundationdb-recordlayer}.
The \emph{server} thus appears to provide strong transactional semantics. The failures
occur at the \emph{client}, where the sync daemon projects this rich distributed state
onto a local filesystem that cannot represent it. The irony is structural: ACID
guarantees on the server are erased by the projection onto the client.

\subsection{The \texttt{fileproviderd} Daemon}

Since macOS High Sierra, the actual work of iCloud Drive synchronization is performed
by \texttt{fileproviderd}, a system daemon that manages the File Provider framework
\cite{borrill2025storage-sync}. This daemon handles metadata synchronization,
on-demand file access through placeholders, and conflict resolution for both iCloud
and third-party providers (Dropbox, OneDrive).

The daemon exhibits several documented failure modes under load:
\begin{itemize}[nosep]
  \item Files hang indefinitely in ``Uploading'' state during large sync operations,
  \item CPU and memory consumption spikes during bulk synchronization,
  \item Error propagation from CloudKit is incomplete---applications receive no
        notification when server-side operations fail,
  \item Rapid multi-device edits produce merge conflicts that the daemon resolves
        by creating duplicates or silently discarding one version.
\end{itemize}

The daemon's behavior can be observed via
\texttt{log stream --predicate 'subsystem == "com.apple.FileProvider"'}, but the
logs reveal an eventual-consistency model with no mechanism for applications to
verify that their writes have been durably committed to the cloud. The daemon
operates as a best-effort system that presents itself as a reliable one.

\subsection{Dataless Files and the Eviction Problem}

Since macOS Sonoma, iCloud Drive uses APFS \emph{dataless files}: file metadata and
extended attributes are stored locally, but data extents are absent
\cite{eclectic-dataless}. A dataless file:
\begin{itemize}[nosep]
  \item is listed by \texttt{ls} and returns valid metadata from \texttt{stat()},
  \item has no data blocks allocated,
  \item triggers an implicit network fetch when read,
  \item is indistinguishable from a real file without inspecting the \texttt{SF\_DATALESS} flag.
\end{itemize}

This breaks the expectation---implicit in decades of POSIX practice, if not in the
formal specification---that \emph{listed files are present and readable}. Tools that
rely on this expectation---including \texttt{git}, \texttt{make}, \texttt{xcodebuild},
and AI coding agents---fail unpredictably.

\section{Incompatibility 1: Time Machine}
\label{sec:timemachine}

Time Machine archives filesystem trees indexed by time. iCloud Drive negotiates
distributed event graphs. A tree snapshot is not equivalent to a distributed agreement
state.

\subsection{The Projection}

Time Machine defines an implicit projection:
\[
F: \text{Sync Category} \rightarrow \text{Snapshot Category}
\]

It forgets the event graph and retains only the instantaneous local tree. This
projection is not faithful: distinct distributed histories can map to the same local
snapshot; the same distributed history can map to different snapshots depending on
timing \cite{borrill2026filesync}.

\subsection{What Is Lost}
\begin{itemize}[nosep]
  \item Concurrency lineage
  \item Device-of-origin
  \item Conflict sets
  \item Merge provenance
  \item Hydration status (dataless vs.\ materialized)
\end{itemize}

\subsection{The Sonoma Cascade}

macOS Sonoma demonstrated this incompatibility dramatically: multiple users reported
Time Machine backups failing to complete while iCloud Drive was actively syncing.
The reported root cause was a circular dependency: Time Machine requires a stable
filesystem state to snapshot; iCloud's sync daemon continuously modifies the
filesystem; the FSEvents database does not settle; and Time Machine cannot acquire
a coherent change set \cite{eclectic-sonoma-tm}.

\subsection{Restore Semantics}

When a Time Machine backup is restored while iCloud is active, the sync daemon
may treat cloud state as authoritative and overwrite the restored local state
\cite{apple-tm-restore}. The ``correct past'' that Time Machine preserved can be
destroyed by the subsequent sync. This is not a bug in either system. It is a
structural consequence of two systems with incompatible models of truth.

\section{Incompatibility 2: Git}
\label{sec:git}

Git depends on POSIX filesystem semantics at a level of detail that iCloud cannot
provide.

\subsection{Git's Locking Model}

Git implements mutual exclusion through lock files created with
\texttt{open(O\_CREAT|O\_EXCL)}---an atomic create-if-not-exists operation. The
write-sync-rename pattern (\texttt{write} $\to$ \texttt{fsync} $\to$ \texttt{rename})
provides transactional semantics: either the update completes fully, or the repository
is unchanged \cite{git-lockfile}.

\subsection{How iCloud Defeats Git}

\begin{enumerate}[nosep]
  \item \textbf{Lock file propagation}: iCloud treats \texttt{.git/index.lock} as an
        ordinary file and syncs it to other devices. A device receiving this file
        believes another process holds the lock and refuses to operate.
  \item \textbf{Ref corruption via numbered suffixes}: When iCloud detects concurrent
        modifications to \texttt{refs/heads/main}, it renames one version to
        \texttt{refs/heads/main 2}. Git never looks for numbered suffixes. The commit
        is orphaned and invisible.
  \item \textbf{Packfile tearing}: Because iCloud syncs files independently,
        a \texttt{.pack} file and its corresponding \texttt{.idx} file can arrive
        at a remote device in inconsistent states. If the packfile is still being
        written when sync begins, the result is an index whose offsets no longer
        match the pack contents.
  \item \textbf{Race conditions}: Git's \texttt{rename()} of \texttt{index.lock} to
        \texttt{index} races with iCloud's sync of the lock file itself. The sync
        daemon can observe and propagate intermediate states that git intends to be
        invisible.
\end{enumerate}

\subsection{Observed Failures}

These failure modes are not hypothetical. Multiple instances of iCloud-induced git
corruption have been directly observed during the preparation of this document and
related work, including: iCloud propagating intermediate file states during LaTeX
collaboration; silent filename swapping that undid deliberate file creation; and
the sync daemon seizing \texttt{.git/index.lock} within seconds of
\texttt{git init}, rendering a repository permanently inoperable with ``Operation
not permitted'' on every file. These incidents are documented with verbatim error
output in Appendix~\ref{sec:appendix-failures}.

\subsection{The Bundle Workaround}

The established workaround is to store git repositories outside iCloud and export
\texttt{.bundle} files for archival \cite{borrill2026wdcgw-session}. A bundle is a
single file; iCloud's conflict resolution (numbered suffixes) produces two complete
bundles rather than a corrupted repository tree. The workaround succeeds precisely
because it \emph{removes git from the cloud sync system entirely}. This is an
admission that the two systems are structurally incompatible.

\section{Incompatibility 3: Automated Toolchains}
\label{sec:toolchains}

Any automated toolchain---build systems, continuous integration pipelines,
scripting frameworks, coding agents---assumes it is operating on a local POSIX
filesystem with strong consistency, immediate write visibility, and no external
modification of files between operations. The following failure modes are logical
consequences of composing such a toolchain with iCloud's sync semantics. Some have
been directly observed; others follow from the architectural mismatch documented
in the preceding sections.

\subsection{The Toolchain-Sync Race}

When an automated toolchain edits a file in an iCloud-synced directory, the following race
is structurally possible:
\begin{enumerate}[nosep]
  \item The agent writes an incomplete intermediate state.
  \item The sync daemon detects the change and begins uploading.
  \item The agent completes the edit.
  \item The sync daemon may have already propagated the incomplete state.
\end{enumerate}

The consequence: other devices can receive corrupted files. The agent itself may
read stale or dataless files on subsequent operations.

\subsection{File Watcher Failure}

Claude Code's atomic write operations (write to temporary file, then rename) have
been observed to fail to trigger filesystem event notifications that tools such as
\texttt{watchman} expect \cite{marimo-6784}. Independently, iCloud Drive's own
interaction with file-watcher libraries is known to produce unreliable event
streams \cite{chokidar-881}. The combination means that neither the agent nor
the sync daemon can maintain reliable knowledge of the filesystem's current state.

\subsection{The Placeholder Trap}

When ``Optimize Mac Storage'' is enabled, files that an agent successfully read
in one session may be evicted by iCloud between sessions. On the next invocation,
the agent encounters dataless files: \texttt{stat()} succeeds, but \texttt{read()}
triggers a network fetch that may time out, return stale data, or fail entirely.
The filesystem has changed state \emph{underneath the agent without any operation
by the agent}. This is a direct consequence of the dataless file architecture
described in Section~\ref{sec:architecture}.

\section{Incompatibility 4: The Filesystem Dialect Problem}
\label{sec:dialect}

Beyond the POSIX violations, iCloud introduces \emph{semantic loss} at the metadata
layer:
\begin{itemize}[nosep]
  \item Extended attributes marked with Apple's \texttt{\#S} suffix are
        preserved during sync; others are silently stripped.
  \item The \texttt{com.apple.ResourceFork} extended attribute is
        typically stripped during sync, even with the \texttt{\#S} suffix
        \cite{eclectic-xattr}.
  \item Case sensitivity, path normalization, and forbidden characters
        differ across platforms (macOS, iOS, Windows via iCloud for Windows),
        creating mismatches invisible at the byte level.
  \item APFS clones---which share data blocks on the originating device---appear as
        full copies on other devices. The clone semantics do not replicate.
\end{itemize}

These losses are individually minor but cumulatively devastating. A project that
depends on extended attributes, resource forks, or case-sensitive paths will
experience silent corruption that no error message reveals.

\section{Case Study: The 366\,GB Archive}
\label{sec:case-study}
\label{sec:archive}

The preceding incompatibilities are not isolated incidents. Over years of use, they
compound into an operational crisis. One of the authors accumulated, through normal
use of iCloud Drive across multiple devices, a 366\,GB ``iCloud Drive (Archive)''
folder containing 110 top-level items that had diverged from the 405\,GB primary
iCloud Drive.

Analysis revealed:
\begin{itemize}[nosep]
  \item 89 folders existed \emph{only} in the archive---files that iCloud had
        silently removed from the primary store.
  \item 57 folders existed in both locations with divergent contents.
  \item Within the shared folders, individual files with identical names had different
        MD5 checksums---silent version drift with no conflict notification.
  \item The FIGURES directory alone contained 1,406 files unique to the archive that
        had vanished from the primary store.
\end{itemize}

Resolving this required writing custom Python scripts to perform MD5-based
comparison of every file in both trees, categorize conflicts, and merge without
data loss \cite{borrill2026archive-merge}. The scripts could not rely on timestamps
(iCloud's conflict resolution had already corrupted them) and instead had to treat
file content as the only source of truth.

The same pattern appeared elsewhere in the filesystem: a personal document
collection had been duplicated into multiple parallel folder hierarchies by iCloud
sync, requiring MD5-based de-duplication to determine which copies were authoritative
\cite{borrill2026collection-manifest}.

This is what the Category Mistake looks like at operational scale. A system that
promises seamless synchronization instead produced hundreds of gigabytes of
unresolvable divergence, silent deletions, and phantom duplicates---requiring
weeks of manual effort and custom tooling to repair.

\section{The Deeper Pattern: FITO in Sync Protocols}
\label{sec:fito}

The preceding four incompatibilities share a common root: all arise from
Forward-In-Time-Only assumptions embedded in iCloud's sync protocol.

\subsection{Timestamp Primacy}

Conflicts are resolved by modification time. This assumes clocks are meaningful
across devices, that ``later'' implies ``more correct,'' and that a single linear
history exists. None of these hold.

\subsection{Intermediate State Leakage}

Application save protocols rely on multi-step sequences intended to be atomic.
iCloud's sync engine propagates intermediate states, exposing partially written
files. This is a direct consequence of replaying a forward timeline rather than
reconciling intended state.

\subsection{Smash-and-Restart Recovery}

When client-side state diverges, iCloud resorts to reset-and-rescan: discard
causal structure, crawl forward again. This approach is fragile, non-idempotent,
and produces the duplicate trees, resurrected deletions, and silent overwrites
that users report.

\subsection{Non-Idempotent Retry Semantics}

Retries duplicate intent rather than reconcile it, producing multiple versions of
what was meant to be a single logical object. This is timeout-and-retry (TAR)
at the filesystem layer---the same pattern that causes retry storms in network
protocols \cite{borrill2025wdcgw}.

\section{Empirical Evidence: Network Partitions Cause Application Failures}
\label{sec:waterloo}

The structural pattern described above---FITO assumptions converting transient
uncertainty into catastrophic, silent, and permanent failures---is not confined
to file synchronization. It has been empirically confirmed across a wide range
of distributed systems by the University of Waterloo group.

Alquraan et al.\ conducted a comprehensive study of 136 system failures
attributed to network-partitioning faults across 25 widely used distributed
systems \cite{alquraan2018analysis}. Their findings are striking:
\begin{itemize}[nosep]
  \item \textbf{80\%} of the failures had catastrophic impact on the system.
  \item \textbf{Data loss was the most common consequence} (26.6\% of failures),
        followed by reappearance of deleted data, broken locks, and system crashes.
  \item \textbf{90\%} of the failures were \emph{silent}---the system did not
        raise an alert, write a log entry, or return an error to the client.
  \item \textbf{21\%} led to \emph{permanent damage} that persisted even after
        the network partition healed.
  \item The majority of failures required little or no client input, could be
        triggered by isolating a single node, and were deterministic.
\end{itemize}

The extended study by Alkhatib et al.\ on \emph{partial} network
partitions---where some but not all nodes lose connectivity---found the same
pattern in 13 popular systems \cite{alkhatib2023partial}. The failures were
catastrophic, easily manifested, and mainly due to design flaws in core system
mechanisms including scheduling, membership management, and ZooKeeper-based
configuration management.

These results are directly relevant to iCloud Drive. iCloud operates in exactly
the regime these studies characterize: intermittent connectivity between devices
(a laptop lid closes, a phone enters a tunnel, an airplane enters flight mode)
creates network partitions that the sync protocol must handle. The Waterloo
findings confirm that the consequences are not merely theoretical. Silent data
loss, permanent damage, and undetected divergence are the \emph{empirically
measured} outcomes of partition-handling failures across the industry's most
widely deployed systems.

\subsection{The Limits of Overlay Solutions}

The Waterloo group proposed Nifty, a transparent overlay network layer that masks
partial partitions by rerouting packets around failed links
\cite{alkhatib2023partial}. While the empirical diagnosis is precise, the
proposed solution illustrates a characteristic limitation: it operates above the
link layer (L3/L4 overlay), attempting to compensate for failures that originate
at or below L2.

This is the same architectural pattern that produces iCloud's failures: applying
higher-layer workarounds to lower-layer problems. Overlay rerouting cannot
address the fundamental issue---that the link itself has entered an ambiguous
state where delivery is unknown---because the overlay has no access to the
bilateral link state. The overlay can detect that packets are not arriving and
reroute them, but it cannot determine whether the original packets were delivered,
partially delivered, or lost. It therefore introduces the same three-state
ambiguity (delivered, not delivered, unknown) that FITO protocols collapse into
error.

As demonstrated by the OAE link protocol (Section~\ref{sec:oae}), resolving
partition-induced ambiguity requires bilateral state at the link boundary itself.
The fix must be at L2, not above it, because that is where the causal information
about delivery exists. No amount of overlay routing can recover information that
was never captured at the point of uncertainty.

\subsection{The Broken Link and the Triangle Topology}

There is a deeper objection that must be addressed. If a link is truly broken,
\emph{no information can flow across it in either direction}. Bilateral state
at the endpoints is necessary but not sufficient: both sides know the link is
down (disconnection is the one failure hazard that a direct link can
unambiguously detect \cite{borrill2026rethinking-fabrix}), but neither side
can communicate this knowledge to the other.

This is where the distinction between \emph{links} and \emph{switched networks}
becomes critical. In a switched network, the failure modes are numerous and
opaque: packets are silently dropped, delayed, duplicated, and reordered by
intermediate switches. The endpoints cannot distinguish a slow node from a dead
one, a congested path from a partitioned one. The ambiguity is irreducible
because the intermediate hops destroy causal information.

A direct link has exactly one failure hazard: disconnection. And disconnection
is straightforward to detect---the absence of the expected token is unambiguous.
Both endpoints know the link is down. What they lack is a path to \emph{each
other} through which to coordinate recovery.

The resolution requires a \emph{triangle topology}: a minimum of three nodes
connected by direct links, such that when any single link fails, the remaining
two links provide an alternative path. This is the minimal causal structure in
which shared state can be maintained despite a single-link failure. One round
trip (Alice $\to$ Bob $\to$ Alice) establishes the triangle inequality---the
minimal geometric structure that supports bilateral knowledge
\cite{borrill2026aether}.

This is precisely what overlay networks attempt to provide, but at the wrong
layer. An overlay reroutes packets around a failed \emph{switched} path, but
the rerouted path still traverses switches that can silently corrupt the causal
chain. A triangle of direct OAE links provides the same topological
redundancy with the causal guarantees intact: each link in the triangle
maintains bilateral state, and the failure of any one link is communicated
via the other two with no loss of causal information.

\begin{figure}[ht]
\centering
\includegraphics[width=0.85\columnwidth]{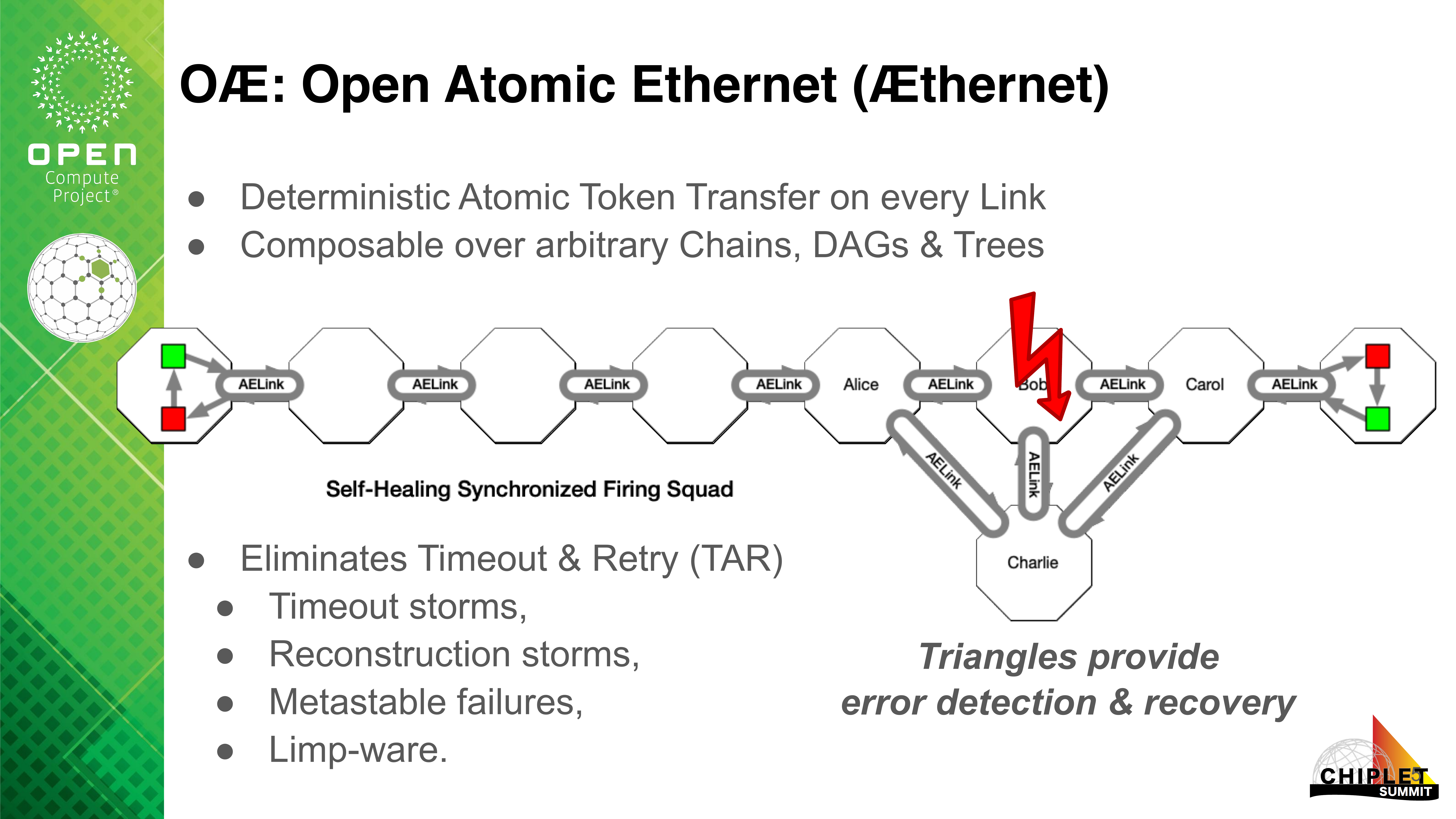}
\caption{A chain of nodes connected by AELinks. Charlie (below) forms a
triangle with Alice and Bob, providing a recovery path when the
Alice--Bob link fails.  From \cite{borrill2026chiplet-summit}.}
\label{fig:triangle-chain}
\end{figure}

\begin{figure}[ht]
\centering
\includegraphics[width=0.85\columnwidth]{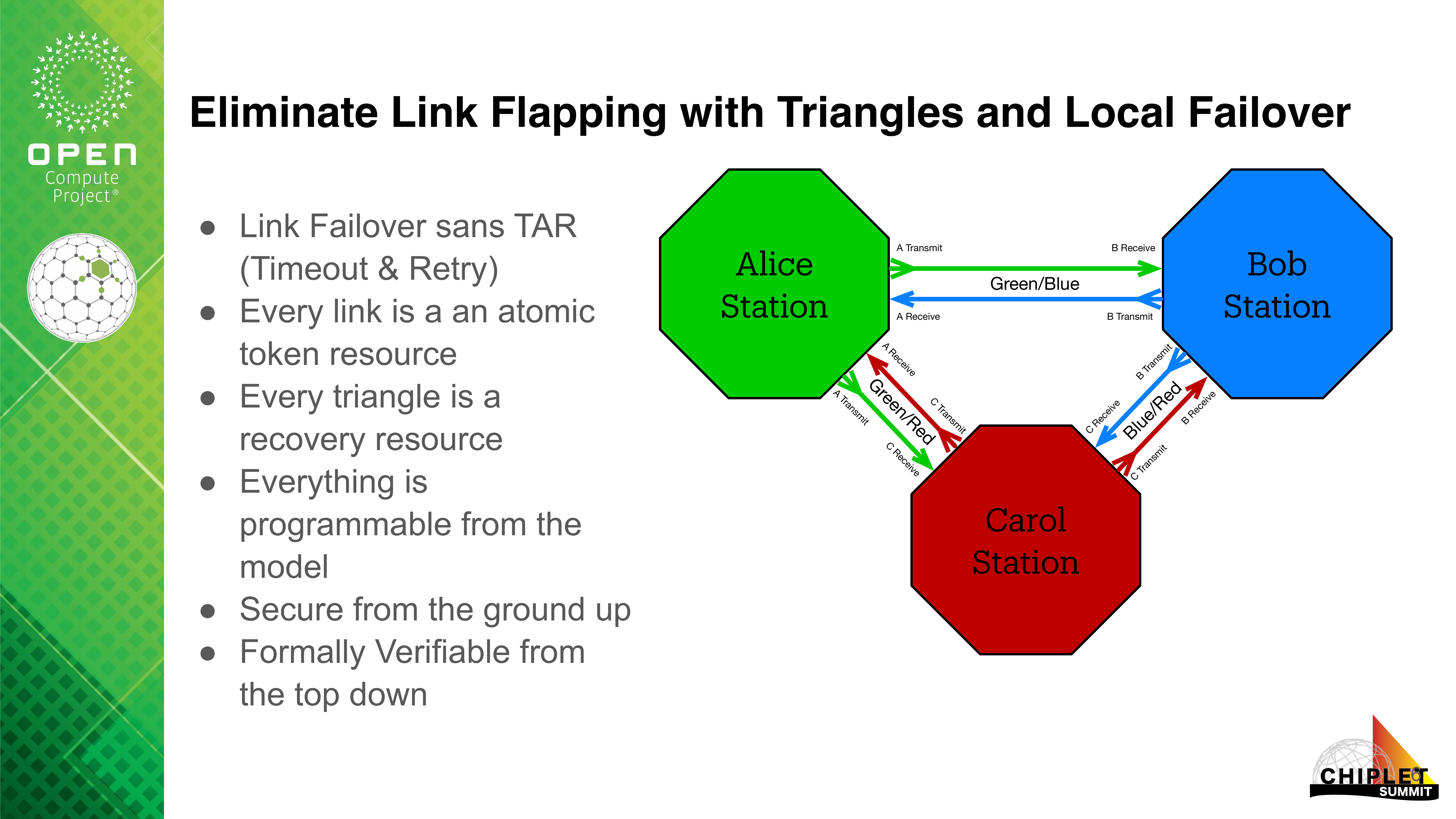}
\caption{Triangle failover as a three-party transaction.  Charlie acts as
the transaction manager (TM) for the Alice--Bob link, coordinating a
2PC recovery via the surviving links.  Every triangle is a recovery
resource; the role of TM rotates symmetrically among all three nodes.
From \cite{borrill2026chiplet-summit}.}
\label{fig:triangle-failover}
\end{figure}

In OAE terminology, each link forms part of a \emph{three-node consensus
group}---a Reliable Link Failure Detector (RLFD) that operates silently at
L2 \cite{borrill2026chiplet-summit}. Every triangle is a recovery resource.
Consider a chain of nodes: Alice--Bob--Carol, connected by direct AELinks
(Figure~\ref{fig:triangle-chain}).  A third node, \emph{Charlie}, sits off
to the side of the chain, with direct links to both Alice and Bob.  When
the Alice--Bob link fails, both endpoints detect the disconnection
unambiguously---and Charlie, as the transaction manager (TM), can
coordinate recovery via its surviving links to each of them.  This is
local failover without timeout-and-retry: the triangle structure provides
the alternative path \emph{and} the causal guarantees simultaneously.

The combinatorial argument makes the advantage concrete. In a switched
network of just four nodes, each pair connected through a Clos fabric,
Colm MacC\'{a}rthaigh has observed that with $n(n-1)/2 = 6$ logical links
and four possible states per link (both directions up, each direction down
independently, both down), the system has $4^6 - 1 = 4{,}095$ possible
failure configurations \cite{borrill2026chiplet-summit}. Link failures in
a Clos are \emph{invisible} to the endpoints---the switch fabric silently
drops, delays, duplicates, or reorders packets without notification. In a
direct-link mesh, every one of those failure modes is \emph{immediately
visible} to the affected endpoints. Triangle supervision eliminates silent
failures.

The triangle failover mechanism is, in fact, a three-party transaction
(Figure~\ref{fig:triangle-failover}).
When the link between Alice and Bob fails, Charlie---positioned off the
chain---acts as the \emph{transaction manager} (TM) and initiates
recovery on the Alice--Bob link by harvesting events from its own ports
to Alice and Bob.  This is exactly the two-phase commit protocol (2PC)
that Jim~Gray described: a coordinator proposes, the participants vote,
and the coordinator resolves \cite{gray1978dbos, gray1993tp}.
The symmetry of the triangle makes the role assignment natural and
complete: Alice is the TM for link Bob--Charlie, Bob is the TM for
link Charlie--Alice, and Charlie is the TM for link Alice--Bob.
Any node that detects a failure on one of its own links initiates
recovery by signalling the TM (the opposite node) to coordinate the 2PC
across the surviving links, using OAE's successively reversible protocols
to return the state machines on both sides to quiescence---ensuring that
no data structures are left stale or inconsistent.  Because each node is
simultaneously a
participant on two links and a transaction manager for the third, the
triangle is the minimal topology in which every link has a designated
recovery coordinator---and that coordinator can always reach both
endpoints, since the two surviving links are, by construction,
operational.

Gray's classical 2PC has a well-known blocking vulnerability: if the
coordinator fails after \textsc{prepare} but before \textsc{commit},
participants are stranded.  In the triangle topology this vulnerability
is eliminated: the coordinator for link Alice--Bob is Charlie, who
is \emph{not on the failed link}.  Charlie's failure would be a
\emph{different} failure---detected and recovered by a different
triangle in the larger mesh.  The failure modes are fully
\emph{separated}: the node that coordinates recovery for a failed link
is, by topological construction, not a victim of that failure.

\subsubsection{Scale Independence without Switches}

\begin{figure}[ht]
\centering
\includegraphics[width=0.85\columnwidth]{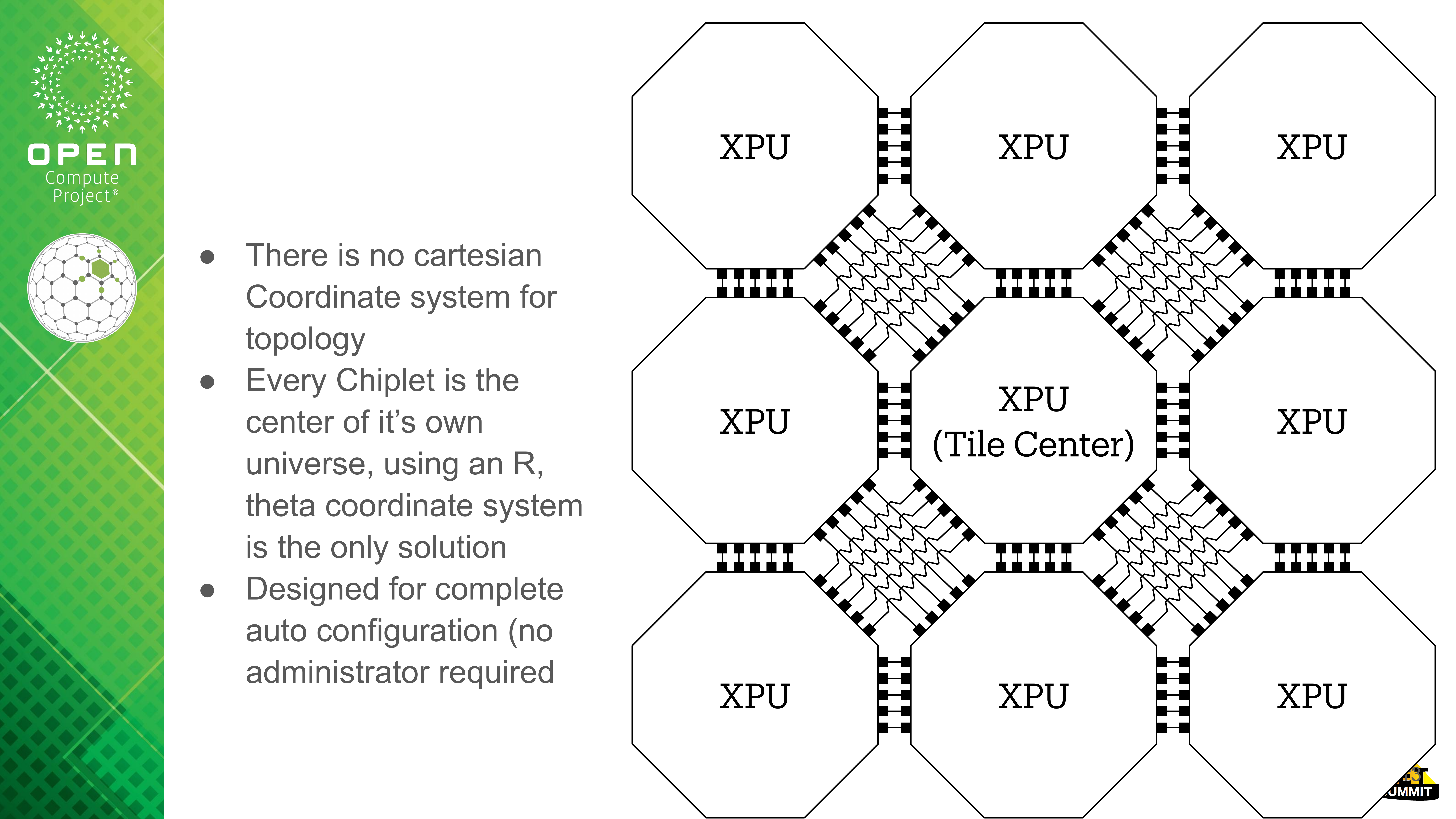}
\caption{A $3 \times 3$ octavalent substrate.  Each XPU connects to its
neighbours in all eight directions (N, S, E, W and all four diagonals)
via direct AELinks.  Every adjacent triple forms a triangle with a
designated TM for each edge.  The array extends indefinitely without
switches, providing scale independence.
From \cite{borrill2026chiplet-summit}.}
\label{fig:xpu-grid}
\end{figure}

The triangle is only a fragment of a larger structure.
Figure~\ref{fig:xpu-grid} shows how triangles compose into an
\emph{octavalent substrate}: a rectangular array of nodes where each
node has eight ports connecting to its neighbours in the four cardinal
directions (N, S, E, W) \emph{and} the four diagonal directions (NE, NW,
SE, SW).  Every adjacent pair of nodes shares a direct link, and every
set of three mutual neighbours forms a triangle with a designated TM for
each edge.

This geometry introduces \emph{scale independence without switches}.
The planar array extends indefinitely in all four cardinal directions;
the diagonal links provide the triangles required for failover at every
point.  Adding a row or column adds nodes and links but introduces no
new switches, no new shared-medium contention points, and no new
categories of silent failure.  The topology scales while preserving the
causal guarantees of every constituent triangle
\cite{borrill2026chiplet-summit}.

Applied to iCloud: the system has no triangle topology at all. Each device
connects to a central cloud server through the public internet---a switched
network with unbounded intermediate hops. There is no direct link between
devices, no bilateral state, no mechanism for one device to learn about
another's state except through the server's eventual-consistency model. The
architecture is structurally incapable of maintaining the shared state that
correct synchronization requires.

But it could be.  Apple users routinely have multiple devices in the same
household---MacBooks, iMacs, iPhones, iPads---all within direct wireless
reach of one another via Wi-Fi, Bluetooth, or even Thunderbolt cable.
These devices already form a physical cluster capable of supporting
triangle topologies.  There is no fundamental reason that household
networks could not adopt OAE's transactional link protocols, enabling
small clusters of Apple devices to maintain bilateral state, detect
failures locally, and recover via triangle failover---all without routing
through a distant cloud server.  The opportunity for Apple to transform
iCloud from an eventually-consistent cloud relay into a resilient local
mesh, augmented by cloud backup, is both technically feasible and
architecturally clear.

\section{Link Flapping and the Destruction of Topology Knowledge}
\label{sec:linkflapping}

The iCloud failure pattern is not confined to file synchronization. The same
structural error---FITO semantics amplifying transient uncertainty into systemic
failure---appears in network fabrics as link flapping.

\subsection{The Scaling Reality}

If a single link exhibits a mean time to flap (MTTF) of $T$, then a system with
$N$ independent links experiences a flap approximately every:
\[
T_{\text{cluster}} \approx \frac{T}{N}
\]

At modern AI scale (hundreds of thousands of optical and electrical links), even rare
transient disturbances become a continuous background process
\cite{borrill2026linkflap}.

\subsection{The Three-State Collapse}

At the moment of disturbance, the system may be in one of three states:
\begin{enumerate}[nosep]
  \item The data was delivered.
  \item The data was not delivered.
  \item The system \emph{cannot currently know} which is true.
\end{enumerate}

FITO protocols collapse all three into a single category: \textbf{error}.
Timeouts, retries, and resets are triggered not by known failure, but by
\emph{unresolved uncertainty}. This is the same epistemic collapse that iCloud
exhibits when it cannot determine which version of a file is authoritative and
resorts to numbered suffixes.

\subsection{Failure Amplification}

The collapse produces a characteristic loop:
\begin{enumerate}[nosep]
  \item Transient physical disturbance introduces uncertainty.
  \item Timeout fires.
  \item Retries increase traffic and noise.
  \item Error rates rise further.
  \item Links retrain or reset.
  \item Global state is disrupted.
\end{enumerate}

Each step is locally rational. The global effect is instability that far exceeds
the underlying physical disturbance. This is structurally identical to iCloud's
failure amplification: a single concurrent edit produces duplicate directory trees,
orphaned refs, and cascading conflicts that far exceed the original divergence.

\subsection{Topology Knowledge Is Information}

When a link flaps, the network loses information about its own topology. Routing
tables, spanning trees, and forwarding state depend on knowledge of which links
are up. A flap forces the system to re-derive this knowledge---but the
re-derivation itself generates traffic, which stresses marginal links, which
causes more flaps. Information is destroyed faster than it can be reconstructed.

This is the same structure as iCloud's FSEvents circular dependency with Time
Machine: the system that tracks state changes is destabilized by the process of
tracking state changes.

\section{The Graph-Theoretic Framework}
\label{sec:graphs}

Distributed systems are trees on top of DAGs on top of graphs
\cite{borrill2026oae-topology}:

\begin{description}[style=unboxed, leftmargin=0pt]
  \item[Graphs:] The physical fabric. Nodes are compute units; edges are
        communication links with asymmetric bandwidth, latency, and failure
        characteristics.
  \item[DAGs:] Causality, scheduling, and locking. Operations must follow
        directed partial orders to preserve consistency.
  \item[Trees:] Namespace hierarchies, leadership structures, commit chains.
        Authority is delegated through tree structures that impose total order
        on specific subsets of the DAG.
\end{description}

iCloud's Category Mistake is the collapse of this layered structure. The sync
engine operates on a DAG (the distributed event graph of uploads, downloads,
merges, and conflict resolutions). Time Machine and git operate on trees (the
filesystem namespace; the commit graph). The filesystem facade pretends to be a
tree when the underlying reality is a DAG. The resulting projection:
\[
F: \text{DAG (distributed sync state)} \rightarrow \text{Tree (local filesystem)}
\]
is not faithful. Information about concurrency, conflict sets, device-of-origin,
and causal lineage is destroyed.

\section{The OAE Resolution}
\label{sec:oae}

Open Atomic Ethernet (OAE) is designed to provide the structural foundation for
resolving the class of failures described above. The resolution is not a ``fix''
for iCloud. It is a demonstration that protocol semantics exist which avoid the
Category Mistake, and that these semantics can be implemented.

\subsection{The Core Principle}

\begin{quote}
\emph{A transfer is not real until it is known to be real.}
\end{quote}

OAE enforces perfect information feedback at the link boundary. Data is only
considered delivered once it has been explicitly reflected back to the sender.
Knowledge, not elapsed time, defines correctness
\cite{borrill2026linkflap}.

\subsection{Bilateral Transactions}

OAE links are not passive channels. They are \emph{joint stateful systems}.
Both peers implement identical state machines that evolve synchronously via the
exchange of fixed-size, causally significant tokens. There is no concept of
master/slave. Either side may assume the role of initiator or responder,
depending on who possesses the token \cite{borrill2026oae-symmetric}.

This is designed to eliminate the class of ambiguity that causes iCloud's failures:
\begin{itemize}[nosep]
  \item No speculative forward progress (no FITO).
  \item No timeout-and-retry (uncertainty pauses progress rather than
        corrupting state).
  \item No epistemic collapse (the three states---delivered, not delivered,
        unknown---remain distinguished).
\end{itemize}

\subsection{Symmetric Reversibility}

For every operation on an OAE link, there exists a logically defined inverse that
restores the prior state \cite{borrill2026oae-reversible}:
\begin{itemize}[nosep]
  \item Partial transactions can be aborted cleanly, returning to equilibrium
        with no partial state leaked on either side.
  \item Errors (bit flips, packet loss) can be rolled back without corrupting
        state.
  \item All token transfers are atomic: they either complete fully or leave the
        system unchanged.
\end{itemize}

Applied to file synchronization, this means: a sync operation that encounters
uncertainty does not produce a numbered suffix, a duplicate directory, or a
silently overwritten file. It \emph{reverses} to the last known-good state and
waits for resolution. Recovery preserves conserved invariants rather than
replaying timestamps.

\subsection{Conservation of Information}

OAE treats every token as a conserved object: inserted, tracked, and retired
without ambiguity. No token is ever ``lost''---it is either delivered or
explicitly rejected. This is the Conservation Principle applied to networking:
the total information in the system is accounted for at all times.

Contrast with iCloud: when a file is modified concurrently on two devices, the
information about \emph{both} modifications must be preserved. iCloud's
last-writer-wins destroys the losing modification. OAE's bilateral protocol
ensures that both sides \emph{know} about the divergence before either side
commits, enabling semantic reconciliation rather than timestamp arbitration.

\subsection{Surviving Link Flapping}

Because OAE links maintain bilateral state, a transient disturbance does not
destroy topology knowledge. When a link flaps:
\begin{enumerate}[nosep]
  \item Both sides detect the disturbance through the absence of the expected
        token.
  \item The link enters a recovery state (not an error state).
  \item The bilateral state machine determines what was in flight at the moment
        of disturbance.
  \item Any in-flight tokens are either confirmed delivered or reversed.
  \item Normal operation resumes from a \emph{known} state, not a \emph{guessed}
        state.
\end{enumerate}

Under this model, the failure is contained within the link's own failure domain.
No routing table invalidation is required. No spanning tree recalculation is
triggered. No retry storm propagates. These are design properties of the OAE
link protocol; formal proofs and quantitative validation are subjects of ongoing
work.

\section{Implications for a Correct Sync Model}
\label{sec:implications}

The preceding analysis implies requirements for any system that claims to
synchronize files correctly across distributed devices:

\begin{enumerate}[nosep]
  \item \textbf{Correctness over state, not event history.}
        Reconciliation must operate on invariants, not timestamps.
  \item \textbf{Application-level atomicity preserved end-to-end.}
        Multi-file save sequences must not leak intermediate states.
  \item \textbf{Concurrency explicit and semantically handled.}
        Concurrent modifications must be surfaced as a structured divergence,
        not arbitrated by clocks.
  \item \textbf{Recovery by invariant, not by timestamp.}
        Restoring ``the state at time $t$'' is insufficient; the system must
        restore conserved invariants.
  \item \textbf{Audit and rollback via causal structure.}
        The event graph must be preserved, not projected onto a linear chain.
  \item \textbf{Bilateral confirmation before commitment.}
        No modification is considered ``synced'' until both sides confirm
        receipt and consistency.
  \item \textbf{Conservation of information.}
        Every file version, every modification, every conflict must be
        accounted for. Silent destruction of data is a protocol violation.
\end{enumerate}

These are not aspirational. They are the properties that OAE is designed to
provide at the link layer. Extending them to file synchronization is an
engineering problem, not a theoretical impossibility.

\section{Summary}

iCloud Drive fails---with Time Machine, with git, with Claude Code, with any tool
that assumes POSIX filesystem semantics---because it commits the same Category
Mistake that pervades distributed computing: projecting a distributed causal graph
onto a linear temporal chain and treating the projection as authoritative.

The failures are not random bugs. They are projection artifacts---predicted by
Parker et al.\ in 1983 and confirmed by four decades of operational experience:
\begin{itemize}[nosep]
  \item Time Machine captures a snapshot of a moving target and calls it history.
  \item Git's lock files become propagation vectors for corruption.
  \item Automated toolchains read dataless files that \texttt{stat()} says exist
        but contain no data.
  \item Extended attributes are silently stripped, destroying metadata invariants.
  \item Link flapping amplifies transient uncertainty into systemic collapse.
  \item Years of normal use produce hundreds of gigabytes of unresolvable divergence,
        requiring custom tooling to repair.
\end{itemize}

Open Atomic Ethernet is designed to demonstrate that these failures are not
inevitable. By replacing FITO semantics with bilateral, reversible,
conservation-preserving transactions, OAE aims to align protocol behavior with
physical reality: uncertainty pauses progress rather than corrupting state;
recovery restores known invariants rather than replaying guessed timelines;
information is conserved, not destroyed.

The question is not whether iCloud can be fixed. The question is whether we are
willing to recognize that the Category Mistake---treating cloud sync as a
filesystem, treating partial order as total order, treating uncertainty as
error---is the source of the problem, and that the solution has been available
all along in the structure of physics itself.

\section*{Acknowledgements}

\subsection*{The Mulligan Stew Gang}

The ideas in this document were sharpened over 130+ consecutive weeks of Friday
morning sessions with the Mulligan Stew Gang, a working group that evolved from
ItsAboutTime.club. The members---physicists, mathematicians, computer scientists,
network architects, and practicing engineers---wish to remain anonymous.



\appendix
\section{Observed Failure Log}
\label{sec:appendix-failures}

This appendix documents specific iCloud Drive failures observed during the
preparation of this document and related work. Each incident is recorded with
dates, error output, and the resolution applied. The intent is to provide a
growing evidence base that can be extended as new failures are encountered.

\medskip
\noindent\textbf{Summary of Incidents.}
Table~\ref{tab:incident-summary} provides a condensed view of all documented
incidents, their failure mechanisms, and data-loss risk.

\begin{table}[ht]
\centering
\small
\begin{tabular}{clllc}
\toprule
\textbf{\#} & \textbf{Incident} & \textbf{Mechanism} & \textbf{Failure class} & \textbf{Data loss} \\
\midrule
1 & LaTeX+Git corruption    & Intermediate state propagation & Sync race         & Yes \\
2 & Silent filename swap    & Conflict mis-resolution        & Metadata mutation  & Potential \\
3 & Repository seizure      & Daemon exclusive lock           & Permission denial  & No (blocked) \\
4 & Keynote image damage    & Package non-atomicity           & Bundle tearing     & Yes \\
5 & Keynote stuck on open   & Dataless file stall             & Eviction           & No (blocked) \\
6 & Shared folder deletion  & Silent local removal            & State divergence   & Yes (confirmed) \\
7 & Integrity audit failure & On-demand eviction mid-session  & Eviction + lock    & Potential \\
8 & AI-generated doc conflict & Concurrent VM/daemon write     & Sync race (LWW) & Potential \\
\bottomrule
\end{tabular}
\caption{Summary of observed iCloud Drive failure incidents.
``Data loss'' indicates whether user data was destroyed or at risk.
All incidents were observed during normal single-user operation with no
concurrent access from other devices.}
\label{tab:incident-summary}
\end{table}

\subsection{Incident 1: LaTeX + Git Collaboration Corruption}
\label{sec:incident-latex-git}

\textbf{Date}: 02026-FEB-22. \textbf{Context}: Collaborative LaTeX manuscript
tracked in Git, stored in iCloud Drive.

During multi-author editing, iCloud's sync daemon propagated intermediate file
states to other devices. Figure paths became invalid as iCloud renamed files
during sync. Git operations failed with spurious merge conflicts that had no
basis in actual concurrent edits \cite{borrill2026latex-git}.

\textbf{Resolution}: The entire repository was moved out of iCloud Drive to a
local directory. Collaboration continued via Git remote, with no further
corruption.

\subsection{Incident 2: Silent Filename Swap}
\label{sec:incident-filename-swap}

\textbf{Date}: 02023-MAY (approx). \textbf{Context}: Creating a new LaTeX
document by copying an existing file in an iCloud-synced directory.

iCloud silently swapped filenames during a copy operation. A LaTeX source
comment records the event directly: ``I edited Subtime-copy.tex to create
this---iCloud should leave it alone and not swap the filenames back''
\cite{borrill2023slowdown}. The sync daemon treated a deliberate file creation
as a conflict requiring resolution, and ``resolved'' it by undoing the user's
work.

\textbf{Resolution}: Manual correction. The comment was left in the source as
a warning.

\subsection{Incident 3: Repository Seizure During \texttt{git init}}
\label{sec:incident-git-init}

\textbf{Date}: 02026-FEB-22. \textbf{Context}: Attempting to create a Git
repository in the iCloud-synced directory containing this paper's source files
(four files, under 200\,KB total, no concurrent access from any other device).

When \texttt{git init} was run, the sync daemon immediately seized the newly
created \texttt{.git/} directory. The first \texttt{git commit} failed:

\begin{verbatim}
fatal: Unable to create '.git/index.lock': File exists.
\end{verbatim}

\noindent Attempting to remove the stale lock file:

\begin{verbatim}
rm: cannot remove '.git/index.lock': Operation not permitted
\end{verbatim}

\noindent Attempting to remove the entire \texttt{.git/} directory:

\begin{verbatim}
rm: cannot remove '.git/index.lock': Operation not permitted
rm: cannot remove '.git/config': Operation not permitted
rm: cannot remove '.git/HEAD': Operation not permitted
rm: cannot remove '.git/hooks/commit-msg.sample': Operation not permitted
rm: cannot remove '.git/hooks/pre-commit.sample': Operation not permitted
rm: cannot remove '.git/refs/heads': Operation not permitted
rm: cannot remove '.git/refs/tags': Operation not permitted
    [... 24 files total, every one "Operation not permitted"]
\end{verbatim}

\noindent The sync daemon had acquired exclusive access to every file in the
\texttt{.git/} directory and would not release any of them. The repository was
permanently inoperable within seconds of its creation.

\textbf{Resolution}: The repository was initialized outside iCloud Drive, source
files were copied in, and the resulting \texttt{.bundle} file was placed in the
iCloud-synced directory for archival---exactly the workaround documented in
Section~\ref{sec:git}. The paper that documents the incompatibility could only
be version-controlled by routing around it. The trapped \texttt{.git/} directory
had to be deleted manually from Finder on the host Mac.

\subsection{Incident 4: Keynote Package Image Degradation}
\label{sec:incident-keynote}

\textbf{Date}: 02026-FEB-22. \textbf{Context}: Opening a Keynote presentation
stored in iCloud Drive.

Upon opening \texttt{DAE-Investor-Primary copy.key}---note the ``copy'' suffix
indicating iCloud had previously created a conflict duplicate---Keynote displayed
a warning dialog:

\begin{quote}
\emph{``This presentation contains images that may be damaged and appear in
low resolution. Would you like to open the file anyway?''}
\end{quote}

Keynote files are macOS package bundles: directories containing XML metadata,
theme data, and embedded media files in an internal structure with cross-references
between components. iCloud syncs the component files independently rather than
treating the package as an atomic unit. When timing or fidelity of the sync breaks
the internal consistency of the package, embedded resources---in this case,
presentation images---are degraded or lost.

This failure demonstrates a distinct incompatibility axis from the git and
filesystem failures: application bundle formats that depend on internal structural
consistency are vulnerable to iCloud's file-by-file sync model. The same class of
failure can affect any macOS package format, including \texttt{.pages},
\texttt{.numbers}, Xcode projects (\texttt{.xcodeproj}), and Core Data stores.

\textbf{Resolution}: No automated resolution available. The damaged images would
need to be re-inserted manually from original sources, if they still exist
elsewhere.

\subsection{Incident 5: Keynote File Stuck on Open}
\label{sec:incident-keynote-stuck}

\textbf{Date}: 02026-FEB-22. \textbf{Context}: Attempting to open a Keynote
presentation stored in iCloud Drive in order to export it as PDF.

While attempting to open \texttt{Spec 0.7A release.key} from the
\texttt{PRESENTATIONS/OAE-Presentations/} directory, Keynote displayed a
progress bar that froze indefinitely. The file could not be opened or exported.
This occurred immediately after Incident~4 (image degradation in a different
Keynote file), suggesting that iCloud's sync state for Keynote package bundles
was in a degraded condition.

This failure demonstrates the \emph{dataless file} problem at the application
level: the file metadata was present (Finder listed it, Keynote attempted to
open it), but the actual data extents---the internal package contents---were
either absent, partially hydrated, or corrupted during on-demand download.
Keynote's progress bar represented the application waiting for data that iCloud
could not deliver.

\textbf{Resolution}: None during the session. The file remained inaccessible.

\subsection{Incident 6: Silent Shared Folder Deletion}
\label{sec:incident-shared-folder}

\textbf{Date}: 02023-APR-05. \textbf{Context}: A shared folder
(\texttt{@DAE-TEAM}) used for collaboration, stored in iCloud Drive on a
Mac (``BozProg Monster'').

\begin{figure}[ht]
\centering
\includegraphics[width=0.85\textwidth]{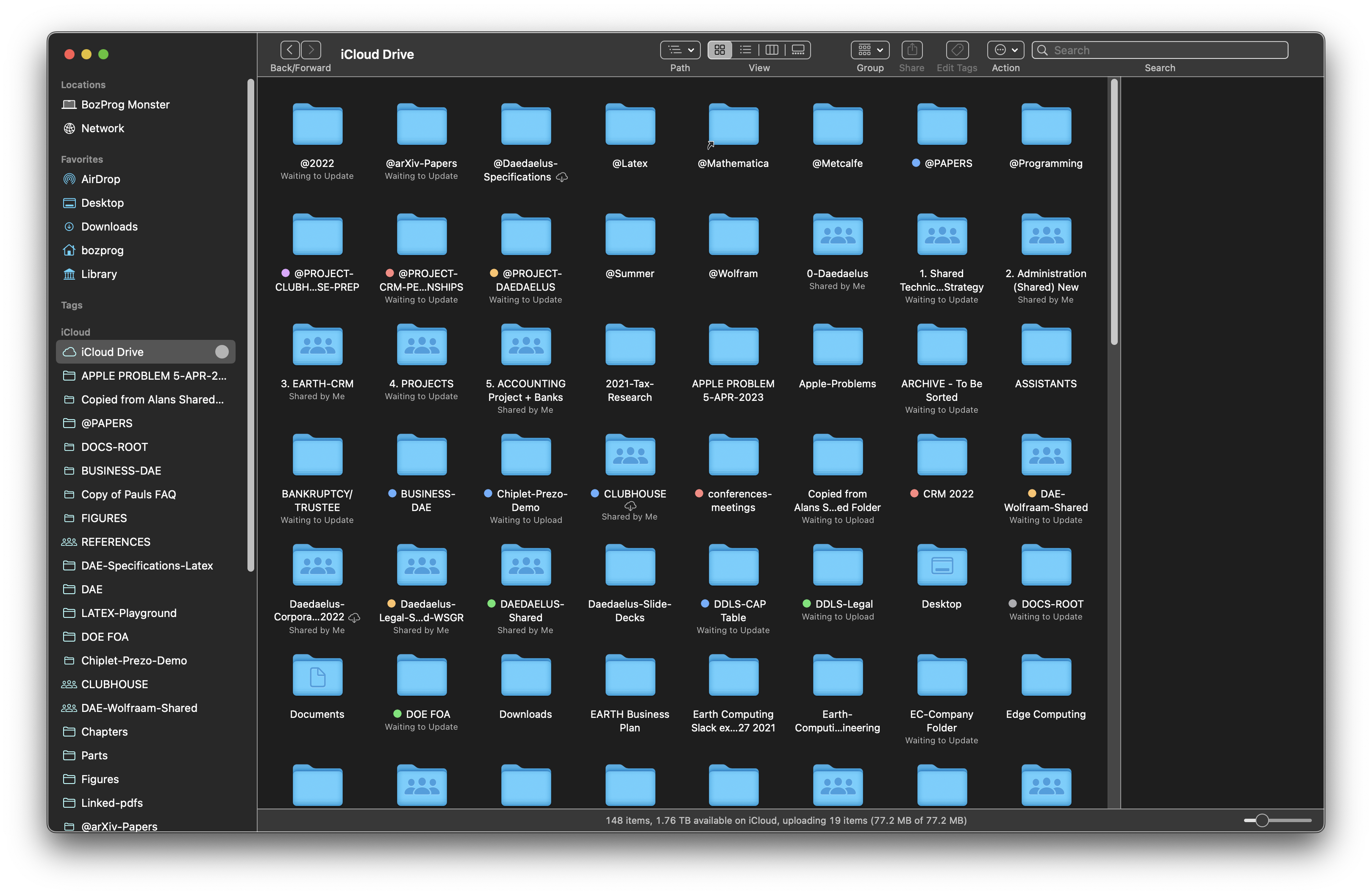}
\caption{Finder icon view, 02023-APR-05 09:08.  The \texttt{@DAE-TEAM} shared
folder is absent from the iCloud Drive listing.  No error or notification
is displayed.}
\label{fig:incident6-icon}
\end{figure}

\begin{figure}[ht]
\centering
\includegraphics[width=0.85\textwidth]{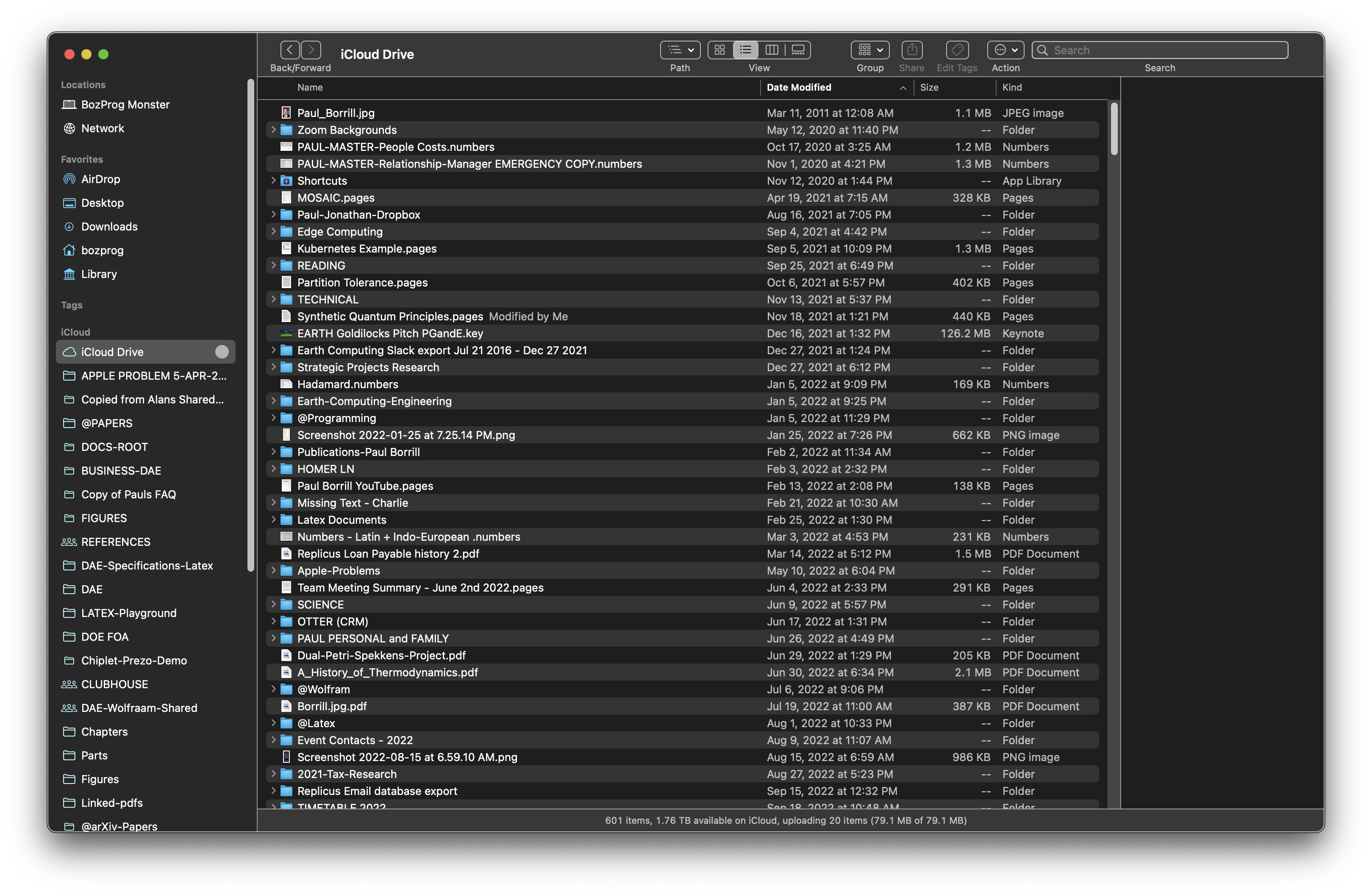}
\caption{Finder list view, same time.  The folder does not appear in
any view mode, confirming its complete absence from the local filesystem.}
\label{fig:incident6-list}
\end{figure}

\begin{figure}[ht]
\centering
\includegraphics[width=0.85\textwidth]{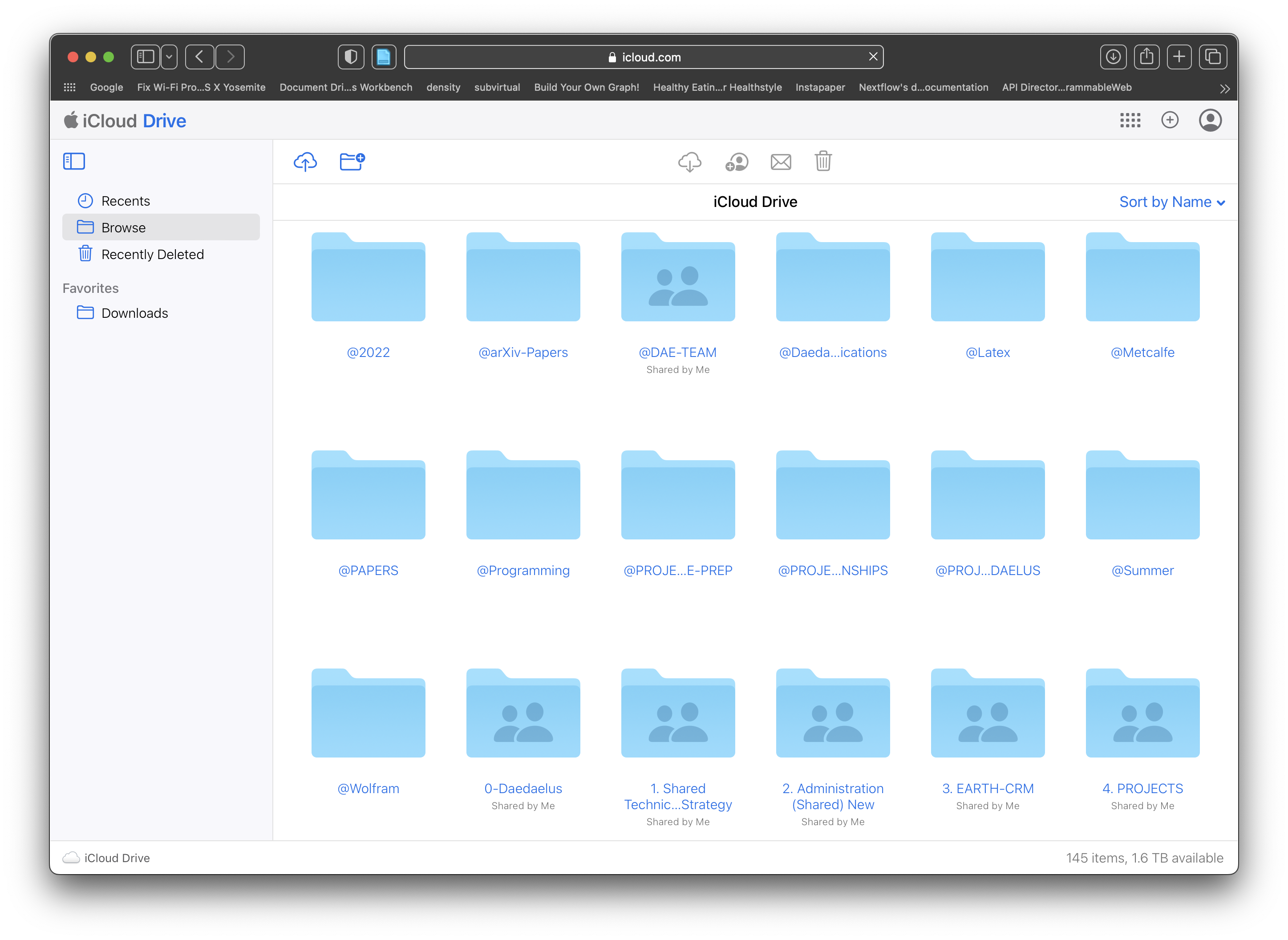}
\caption{iCloud.com (Safari), 02023-APR-05 09:27.  The \texttt{@DAE-TEAM}
folder is visible on the server, marked ``Shared by Me.''  The server and
client have divergent views of the same directory tree.}
\label{fig:incident6-web}
\end{figure}

The folder \texttt{@DAE-TEAM}, which had been shared with team members and was
in active use, disappeared from iCloud Drive on the local machine.  The user
verified the absence systematically:

\begin{enumerate}[nosep]
\item Finder icon view: folder not present (Figure~\ref{fig:incident6-icon}).
\item Finder list view: folder not present (Figure~\ref{fig:incident6-list}).
\item iCloud.com (Safari): folder \emph{is} present, marked ``Shared by Me''
      (Figure~\ref{fig:incident6-web}).
\end{enumerate}

\noindent The folder existed on the server but had been silently removed from
the local filesystem.  There was no error message, no conflict notification,
no entry in Recently Deleted, and no user action that could have caused the
removal.

\begin{figure}[ht]
\centering
\includegraphics[width=0.85\textwidth]{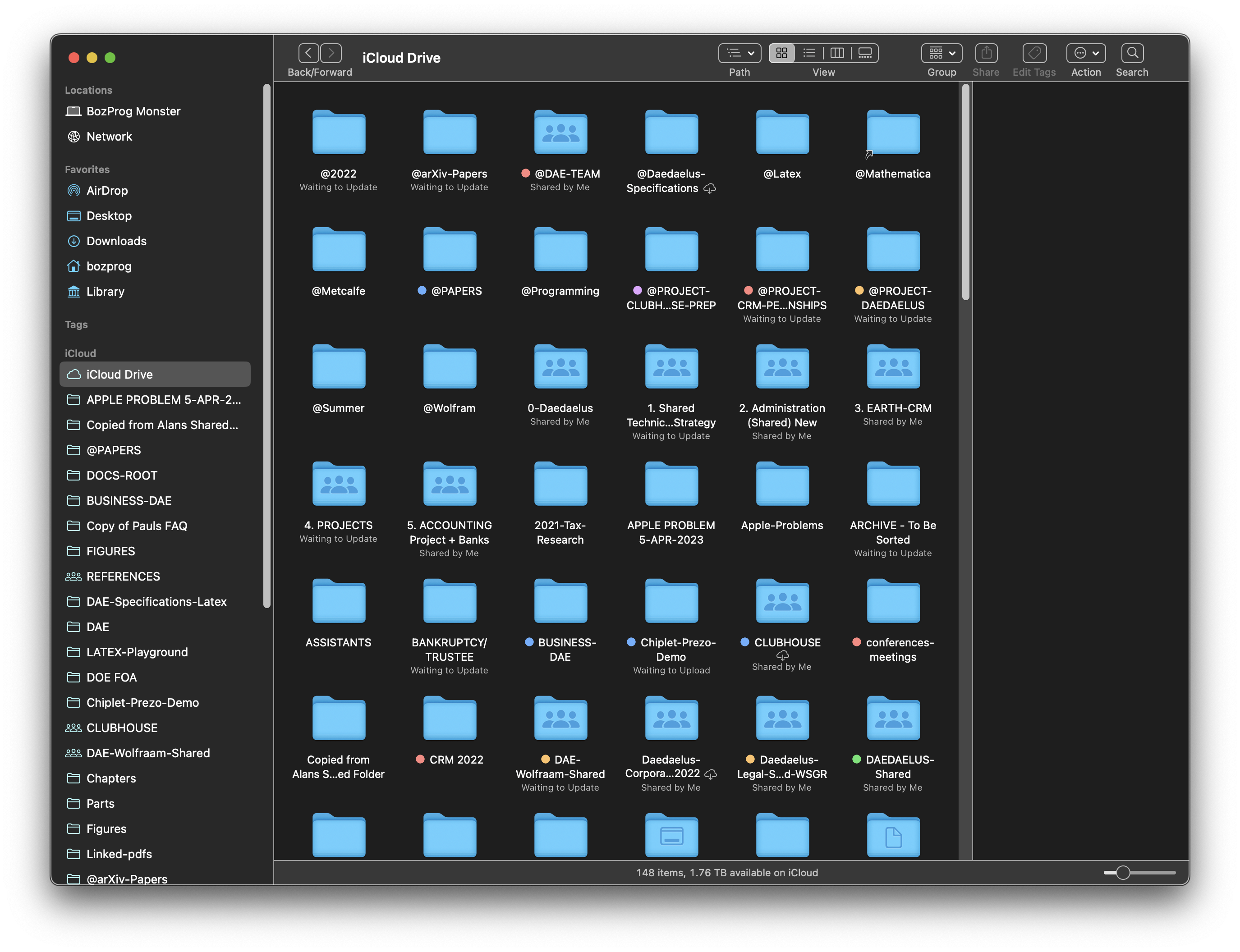}
\caption{Finder after reboot, 02023-APR-05 10:06.  The \texttt{@DAE-TEAM}
folder has reappeared.  The sidebar shows ``Synced with iCloud'' with no
indication that the folder had been missing.}
\label{fig:incident6-restored}
\end{figure}

After a reboot, the folder reappeared locally
(Figure~\ref{fig:incident6-restored}).  iCloud's status indicator showed
``Synced with iCloud---Last sync a moment ago,'' as though nothing had been
wrong.

This incident demonstrates three distinct failures:

\begin{enumerate}[nosep]
\item \textbf{Silent data removal}: a shared folder was deleted from the local
      filesystem with no notification to the user or any participating
      application.
\item \textbf{State divergence}: the server and client had inconsistent views of
      the same directory tree---exactly the condition Parker et al.\ showed is
      unavoidable under network partitioning \cite{parker1983detection}.
\item \textbf{Concealed recovery}: the reboot-triggered resync restored the
      folder, but the system provided no indication that it had ever been
      missing.  From iCloud's perspective, the failure never happened.
\end{enumerate}

\noindent The user documented the incident in a \texttt{.pages} file containing
timestamped screenshots, noting: ``This is not a user error.  This is an iCloud
Error.''

\textbf{Resolution}: The folder was recovered by rebooting the machine, forcing
a full resync.  This is an instance of the ``smash-and-restart'' recovery
pattern described in Section~\ref{sec:fito}: discard local state and rebuild
from the server.  No mechanism existed to diagnose \emph{why} the folder had
vanished, whether any content was lost during the period of absence, or whether
the restored state matched the state before deletion.

\textbf{Postscript} (02026-FEB-26): During a systematic reconciliation of a
366\,GB ``iCloud Drive (Archive)'' against the current iCloud Drive (described in
Section~\ref{sec:case-study}), two files from \texttt{@DAE-TEAM} were found to
exist \emph{only} in the archive---meaning iCloud had at some point silently
deleted them from the primary store.  The reconciliation recovered them.
This confirms that the April 2023 incident was not transient: it resulted in
permanent, undetected data loss that persisted for nearly three years.

\subsection{Incident 7: On-Demand Eviction Prevents File Integrity Audit}
\label{sec:incident-eviction-audit}

\textbf{Date}: 02026-FEB-27. \textbf{Context}: A systematic file integrity
audit of 72 files stored in an iCloud-synced directory
(\texttt{CLAUDE-KEN-BIRMAN/}), performed from a sandboxed Linux VM (Claude
Code session) with read access to the iCloud Drive mount. The purpose was to
compute SHA-256 checksums of every file and verify that each had an identical
copy in a canonical location elsewhere in the same iCloud Drive before
considering the folder for cleanup.

\subsubsection*{Phase 1: Source Checksums (Successful)}

SHA-256 checksums were computed for all 72 files in
\texttt{CLAUDE-KEN-BIRMAN/}. The operation completed without error. All files
were readable and returned stable hashes. This phase took approximately 45
seconds.

\subsubsection*{Phase 2: Canonical Copy Verification (Failed)}

The audit then attempted to checksum the corresponding canonical copies in
four other iCloud-synced directories:

\begin{enumerate}[nosep]
  \item \textbf{CATEGORY-FITO-ANALYSIS/} (40+ subdirectories of researcher
        analyses): Files that had been listed successfully via \texttt{ls}
        earlier in the same session now returned ``No such file or
        directory'' when \texttt{sha256sum} was invoked. iCloud's
        \texttt{fileproviderd} daemon had \emph{evicted} the file contents
        from local cache between the directory listing and the checksum
        operation. The files appeared in directory listings but their data
        extents were no longer materialized on disk.
  \item \textbf{DAE-TECHNICAL-REPORTS/DAE-TR-1006-Birman/}: Every file
        returned ``Permission denied.'' The iCloud sync daemon held
        exclusive access.
  \item \textbf{EMAIL-CONVERSATIONS/}: Pre-existing files (not created
        during this session) had been evicted. Only files \emph{copied
        during this session} remained readable.
  \item \textbf{BIGFAQ/}: Two files were readable but returned
        \emph{different} checksums from the source (see below).
\end{enumerate}

\subsubsection*{Results}

Of 72 files audited:

\begin{center}
\begin{tabular}{llr}
\toprule
\textbf{Category} & \textbf{Reason} & \textbf{Count} \\
\midrule
Verified match     & Copies made during this session & 10 \\
Version mismatch   & Different content at canonical location & 3 \\
Cannot verify      & iCloud eviction or permission denied & 59 \\
\bottomrule
\end{tabular}
\end{center}

\noindent The three version mismatches were:

\begin{center}
\begin{tabular}{p{4.2cm}p{3.8cm}p{4.5cm}}
\toprule
\textbf{Source file} & \textbf{Canonical location} & \textbf{Finding} \\
\midrule
\texttt{faq/FAQ-Birman.tex} \newline (SHA-256: \texttt{c901f6\ldots})
  & \texttt{BIGFAQ/} \newline (\texttt{e28abe\ldots})
  & Different versions \\
\texttt{faq/FAQ-Birman.pdf} \newline (\texttt{3c8549\ldots})
  & \texttt{BIGFAQ/} \newline (\texttt{63ddc6\ldots})
  & Different versions \\
\texttt{leibniz-bridge/}\newline\texttt{Leibniz-Bridge-}\newline\texttt{Synthesis-DRAFT.tex}
  \newline (\texttt{0bf443\ldots}, v0.85)
  & \texttt{CLAUDE-Protocol-}\newline\texttt{Reviews/} \newline (\texttt{836a09\ldots})
  & Source is \emph{newer}; canonical location has older version \\
\bottomrule
\end{tabular}
\end{center}

\noindent The third mismatch is particularly significant: the file being
considered for deletion was the \emph{only} copy of the most recent version.
Had the folder been deleted based on the assumption that canonical copies
existed, the latest version of a 50-page framework document would have been
silently lost.

\subsubsection*{The Temporal Ordering Failure}

The core failure is temporal. The audit followed a straightforward protocol:

\begin{enumerate}[nosep]
  \item List directory contents (\texttt{ls -R}) --- succeeded at time $t_0$.
  \item Compute checksums of source files --- succeeded at time $t_1 > t_0$.
  \item Compute checksums of canonical copies --- \emph{failed} at time
        $t_2 > t_1$.
\end{enumerate}

\noindent Between $t_0$ and $t_2$, iCloud's on-demand eviction daemon
(\texttt{fileproviderd}) offloaded file data from local storage. Files that
were readable at $t_0$ were not readable at $t_2$. The audit assumed---as
any reasonable filesystem client would assume---that a file successfully
listed can be subsequently read. iCloud violated this assumption.

This is the FITO Category Mistake at the filesystem layer: the system
treats a directory listing as a \emph{commitment} that the listed files
are available, but the listing is merely a \emph{snapshot of metadata}
that carries no guarantee about data availability at any future time.
The ``file'' returned by \texttt{ls} is not a file in the POSIX sense;
it is a \emph{promise} that may be revoked without notice.

\subsubsection*{Significance}

This incident is self-referential: the failure occurred during an
audit designed to verify the integrity of files that document the
FITO Category Mistake. The audit could not be completed because the
filesystem committed the same category mistake the files describe.
Specifically:

\begin{enumerate}[nosep]
  \item \textbf{On-demand eviction is invisible}: No error was raised when
        files were evicted. The only indication was a ``No such file or
        directory'' error on a path that \texttt{ls} had listed minutes
        earlier.
  \item \textbf{Permission semantics are non-deterministic}: Files in
        \texttt{DAE-TECHNICAL-REPORTS/} that had been readable in previous
        sessions were locked by the sync daemon during this session.
  \item \textbf{Metadata lies}: \texttt{ls} reports file sizes and
        modification dates for files whose data is not present on disk.
        \texttt{stat()} succeeds on evicted files. Only actual read
        operations reveal the absence.
  \item \textbf{Recovery requires physical access}: The audit could not be
        completed from any sandboxed or remote environment. Full
        verification requires force-downloading all files on the host Mac
        (\texttt{brctl download}) before checksumming---a manual,
        per-directory operation with no batch mode.
\end{enumerate}

\textbf{Resolution}: The folder was \emph{not} deleted. The audit
demonstrated that the information required to make a safe deletion decision
was not obtainable through the filesystem interface that iCloud presents.
The only verified copies were those created during the current session;
all pre-existing copies were either evicted, locked, or mismatched.

\subsection{Incident 8: Real-Time Sync Conflict on AI-Generated Document}
\label{sec:incident-ai-sync}

\textbf{Date}: 02026-FEB-27. \textbf{Context}: A 9-page LaTeX document was
generated by an AI assistant (Claude Opus~4) running in a sandboxed Linux VM
with write access to the user's iCloud Drive. The document---a project plan
for organizing Clubhouse recordings of physics discussions on
causality---was compiled twice with \texttt{pdflatex} (standard practice for
resolving table-of-contents cross-references) and saved to
\texttt{ITSABOUTTIME/@PROJECT-CLUBHOUSE-PREP/}.

Upon opening the resulting PDF in Preview.app, macOS presented the dialog:
``Modifications aren't in sync. Choose which versions to keep,'' offering
two versions:
\begin{enumerate}[nosep]
  \item ``Modified by me'' --- Today, 12:39\,PM
  \item ``Modified on BozProg Monster'' --- Today, 12:39\,PM
\end{enumerate}

Both versions carried identical timestamps. The conflict arose because the
VM-side write (via the mounted iCloud directory) and the
\texttt{fileproviderd} sync daemon each produced a version within the same
second. The LWW protocol could not determine precedence because the
timestamps were indistinguishable---precisely the scenario where
last-writer-wins degenerates into \emph{arbitrary}-writer-wins.

The document being corrupted contained, among other things, a section
recommending that Clubhouse recordings ``be backed up outside iCloud (the
irony of the `Why iCloud Fails' paper is not lost).'' The user captured a
screenshot of the conflict dialog overlaid on the document's table of
contents as primary evidence.

\textbf{Failure mechanism}: Concurrent-write race between a VM filesystem
mount and the iCloud sync daemon, resolved by timestamp comparison on
versions with identical timestamps. This is a degenerate case of LWW
where the ``last'' writer is undefined.

\textbf{Resolution}: The user selected the ``BozProg Monster'' version
(the VM-generated copy), which contained the correctly resolved
table-of-contents cross-references from the second \texttt{pdflatex} pass.
The ``Modified by me'' version was likely the intermediate output from the
first pass, before cross-references were resolved.

\subsection{Incident 9: Asymmetric Write/Delete Permissions in Sandbox Mount}
\label{sec:incident-asymmetric-permissions}

\textbf{Date}: 02026-MAR-04. \textbf{Context}: A file consolidation
operation was performed on iCloud Drive from a Cowork VM (lightweight
Linux sandbox on macOS) with write access to the user's iCloud folder.
The protocol was deliberately designed as a 3-step atomic sequence:
copy, verify (md5sum), then delete originals.

All 123 file copies succeeded (exit code 0). All 123 checksum
verifications passed (zero mismatches). All 17 delete operations
failed uniformly:

\begin{verbatim}
rm: cannot remove '.../@ACTIVE/Metcalfe+Boggs+Borrill/
    Metcalfe+Boggs-content.tex': Operation not permitted
rm: cannot remove '.../@Metcalfe/Metcalfe.bib':
    Operation not permitted
    [... 17 top-level deletes, all denied ...]
\end{verbatim}

\noindent The observed permission model:

\begin{center}
\begin{tabular}{lcc}
\toprule
Operation & Syscall & Result\\
\midrule
Create file & \texttt{open(O\_CREAT)} & Succeeds\\
Write file & \texttt{write()} & Succeeds\\
Create directory & \texttt{mkdir()} & Succeeds\\
Copy file & \texttt{open() + write()} & Succeeds\\
Delete file & \texttt{unlink()} & \textbf{Operation not permitted}\\
Delete directory & \texttt{rmdir()} & \textbf{Operation not permitted}\\
\bottomrule
\end{tabular}
\end{center}

\noindent This violates the POSIX contract for a writable directory:
\[
\mathrm{creat}(dir/f) \nRightarrow \mathrm{unlink}(dir/f)
\]

The asymmetry is structurally isomorphic to the FITO assumption:
forward operations (create, write) succeed; backward operations
(delete, retract) are denied. The sandbox grants monotonic state
growth but forbids retraction of existing state---precisely the
temporal asymmetry that characterizes FITO systems. The consequence
is that \texttt{mv} (rename) is impossible across directory
boundaries, since cross-mount \texttt{mv} is implemented as
\texttt{cp} + \texttt{unlink}, and \texttt{unlink} is denied.

Negative control: the same 3-step protocol on a local (non-iCloud)
APFS filesystem succeeds for all three steps.

\textbf{Resolution}: The system was left with verified duplicates.
The originals persist because the sandbox will not release them.
No data was lost, but the operation could not complete---the
``move'' invariant (exactly one copy) was violated.

Raw logs preserved in \texttt{ETHERNET-50/copy-log.txt},
\texttt{verify-log.txt}, and \texttt{delete-log.txt}.

\subsection{Incident 10: Silent Filename Prefix Corruption}
\label{sec:incident-filename-prefix}

\textbf{Date}: 02026-MAR-07. \textbf{Context}: 17 PDF files were
downloaded from arXiv via Safari and saved directly into
\texttt{CLAUDE-PROJECTS/published/}, a newly created folder on
iCloud Drive. The files were the author's own published papers,
downloaded sequentially over an 18-minute window.

Thirteen files arrived with their original arXiv filenames intact.
Four did not:

\begin{center}
\begin{tabular}{ll}
\toprule
\textbf{Observed filename} & \textbf{Expected filename}\\
\midrule
\texttt{H1CrIO-2602.22350v2.pdf} & \texttt{2602.22350v2.pdf}\\
\texttt{I7uhgv-2603.02603v1.pdf} & \texttt{2603.02603v1.pdf}\\
\texttt{itQDNe-2603.03736v1.pdf} & \texttt{2603.03736v1.pdf}\\
\texttt{nBub83-2602.18723v1.pdf} & \texttt{2602.18723v1.pdf}\\
\bottomrule
\end{tabular}
\end{center}

\noindent Each prefix consists of exactly 6 characters drawn from
\texttt{[A-Za-z0-9]}, followed by a hyphen. Six base64 characters
encode 36 bits of entropy ($\approx 6.87 \times 10^{10}$ possible
values), consistent with a truncated CloudKit record identifier or
\texttt{NSFileCoordinator} disambiguation token.

The prefixes were not present in Safari's download dialog or in the
arXiv source URLs. They appeared only after iCloud Drive's sync
daemon processed the files into the synced folder.

\paragraph{Temporal distribution.}
The corrupted files appeared at positions 5, 7, 11, and 16 in the
17-file download sequence (timestamps 09:34, 09:37, 09:40, 09:48).
There is no temporal clustering---the corruption is non-deterministic.
The same operation (Safari save to iCloud folder) sometimes preserves
the filename and sometimes does not.

\paragraph{Prevalence.}
A systematic scan of the entire iCloud Drive mount identified 17
additional files exhibiting the identical 6-character random prefix
pattern, spanning the \texttt{PAPERS/}, \texttt{STANDALONE/}, and
other folders. The affected files include downloads from arXiv,
ACM Digital Library, ScienceDirect, and Springer---the corruption
is not source-specific.

\paragraph{The smoking gun.}
Two files in \texttt{PAPERS/} are copies of the \emph{same} arXiv
paper (1811.12409v1) but carry \emph{different} random prefixes:

\begin{verbatim}
bHUo8M-1811.12409v1 copy.pdf
q0ZqE1-1811.12409v1.pdf
\end{verbatim}

\noindent This proves the prefix is generated \emph{per download
event}, not per file identity. The same semantic object acquires
different names depending on the timing of the sync layer's
intervention.

\paragraph{FITO interpretation.}
A filename is a semantic identifier---it encodes meaning (here,
the arXiv paper ID, version, and format). The sync layer treats
it as mutable transport-layer metadata that it may modify for
operational convenience (conflict disambiguation). But the filename
is application-layer state that must be preserved. This confusion---treating
semantic content as transport overhead---is precisely the FITO
category mistake: the sync daemon pushes state forward in time,
modifying it unilaterally, without bilateral confirmation that
the modification preserves the user's intent.

Compare with Incident~2 (Section~\ref{sec:incident-filename-swap}):
there, iCloud \emph{swapped} two filenames during a copy operation.
Here, iCloud \emph{prepends} a random token during sync. These are
distinct mechanisms, but they violate the same invariant: the
filename should be preserved as a stable identifier across filesystem
operations. Both arise from the sync layer treating filenames as
state it is entitled to mutate during forward-in-time propagation.

\textbf{Resolution}: The four corrupted filenames were manually
corrected. The 17 additional affected files elsewhere in iCloud
Drive remain uncorrected pending systematic review.

%
%
%
%

\section{Reproducibility Protocol and Negative Control}
\label{sec:appendix-reproducibility}

This appendix documents the experimental environment, audit procedure,
and a negative control test that isolates the observed failures to
iCloud's \texttt{fileproviderd} daemon rather than to the filesystem,
hardware, or toolchain.

\subsection{Environment}
\label{sec:repro-environment}

\begin{description}[style=unboxed, leftmargin=0pt, itemsep=2pt]
  \item[Host machine:] Mac (Apple Silicon), macOS Sequoia 15.x.
  \item[iCloud settings:] ``Optimize Mac Storage'' enabled (default).
        iCloud Drive active, syncing to Apple servers over residential
        broadband.
  \item[Execution environment:] Sandboxed Linux VM (Ubuntu~22.04, ext4
        root filesystem) with read-only FUSE mount of the host's iCloud
        Drive directory.  No write access to iCloud; no direct access to
        \texttt{fileproviderd} or \texttt{brctl}.
  \item[Tools:] GNU \texttt{coreutils} (\texttt{ls}, \texttt{sha256sum}),
        standard POSIX shell.  No iCloud-specific APIs were used.
  \item[Concurrent activity:] No other user or application was modifying
        the audited directories during the test.  A single device was
        active.
\end{description}

\subsection{Audit Procedure}
\label{sec:repro-procedure}

The three-step audit protocol is designed to be trivially reproducible
on any POSIX filesystem:

\begin{enumerate}
  \item \textbf{List} the target directory recursively (\texttt{ls -R}).
        Record the file count and path list.
  \item \textbf{Checksum} every listed file (\texttt{sha256sum}).  Record
        all hashes.
  \item \textbf{Re-checksum} the same files after a delay (minutes to
        hours, depending on session length).  Compare hashes.  Verify
        that every file listed in Step~1 is still readable in Step~3.
\end{enumerate}

\noindent\textbf{Expected result on a conforming POSIX filesystem:}
All files listed in Step~1 remain readable in Step~3. All checksums
match between Steps~2 and~3, unless the user or another process has
modified the files in the interim.

\noindent\textbf{Observed result on iCloud Drive:}
59 of 72 files listed in Step~1 were unreadable in Step~3.
See Incident~7 (Section~\ref{sec:incident-eviction-audit}) for details.

\subsection{Negative Control: Local ext4 Filesystem}
\label{sec:repro-control}

To confirm that the failures are attributable to iCloud and not to the
execution environment, toolchain, or filesystem implementation, the same
three-step protocol was executed on a local ext4 partition within the
same VM session, using files copied from iCloud Drive to local storage.

\begin{description}[style=unboxed, leftmargin=0pt, itemsep=2pt]
  \item[Filesystem:] ext4 (Linux VM root partition).
  \item[Files:] 4 files (73\,KB \texttt{.tex}, 13\,MB \texttt{.pdf},
        16\,KB \texttt{.md}, 8\,KB \texttt{.md}), copied from iCloud
        mount to local directory.
  \item[Procedure:] Steps 1--3 as above, with a 2-second delay between
        Steps~2 and~3.
\end{description}

\noindent\textbf{Result:} All 4 files remained readable across all three
steps.  SHA-256 checksums were identical between Steps~2 and~3.  No
files were evicted, locked, or returned permission errors.  The audit
protocol completed without anomaly.

This confirms that the three-step protocol is sound and that the
failures documented in Incident~7 are specific to iCloud Drive's
on-demand eviction behavior, not to the audit methodology, the VM
environment, or the checksum toolchain.

\subsection{Reproduction Instructions}

To reproduce Incident~7 on a Mac with iCloud Drive and ``Optimize Mac
Storage'' enabled:

\begin{enumerate}
  \item Identify a directory containing files managed by iCloud Drive
        that have not been accessed recently (i.e., likely evicted).
  \item Run \texttt{ls -la} on the directory.  Confirm files are listed
        with sizes and timestamps.
  \item Immediately run \texttt{sha256sum *} on the same directory.
  \item If any file returns ``No such file or directory'' despite being
        listed in Step~2, the eviction failure has been reproduced.
  \item For the permission-denied variant, attempt to read files in a
        directory that \texttt{fileproviderd} is actively syncing.
\end{enumerate}

\noindent\textbf{Note on system logs:} The \texttt{fileproviderd}
daemon's behavior can be observed via:
\begin{verbatim}
log stream --predicate 'subsystem == "com.apple.FileProvider"' \
           --level debug
\end{verbatim}
Correlating log timestamps with eviction events would provide additional
evidence of the causal mechanism.  System logs were not captured during
the Incident~7 session because the audit was performed from a sandboxed
VM without access to the host's log stream.  Future reproductions should
include correlated log output.

\section{Formalization of the Filesystem Contract Violation}
\label{sec:appendix-posix}

The failures documented in this paper can be stated as a violation of
an implicit contract that POSIX-compatible filesystems have honored for
decades.

\subsection{The POSIX Listing Contract (Implicit)}

While the POSIX standard does not formally guarantee that a listed file
will remain readable, the following property has held on every local
filesystem in common use (ext4, XFS, HFS+, APFS, NTFS, ZFS):

\medskip
\noindent\textbf{Property L (Listing Stability):}
\begin{quote}
If \texttt{readdir()} returns an entry $e$ for path $p$ at time $t_0$,
and no process issues \texttt{unlink($p$)} or \texttt{rename($p$, \ldots)}
between $t_0$ and $t_1$, then \texttt{open($p$)} at time $t_1$ succeeds.
\end{quote}

\noindent This property is not merely conventional.  It is a structural
consequence of how local filesystems work: directory entries point to
inodes, and inodes point to data blocks that are allocated on the same
physical medium.  There is no mechanism by which data blocks can
``disappear'' between a directory read and a file open, short of hardware
failure or administrative intervention.

\subsection{iCloud's Violation}

iCloud Drive with ``Optimize Mac Storage'' violates Property~L:

\medskip
\noindent\textbf{iCloud behavior:}
\begin{quote}
\texttt{readdir()} returns an entry $e$ for path $p$ at time $t_0$.
No user process issues \texttt{unlink} or \texttt{rename}.
\texttt{open($p$)} at time $t_1 > t_0$ may fail with \texttt{ENOENT},
because \texttt{fileproviderd} has evicted the data extents between
$t_0$ and $t_1$.
\end{quote}

\noindent The violation is precise: the directory entry persists (metadata
is local), but the data it refers to has been removed by a system daemon
without any user-visible operation.  The file exists in the namespace but
not on the storage medium.  This is not a race condition in the usual
sense---no user process competed for the file.  It is a unilateral
revocation of data availability by the sync infrastructure.

\subsection{Consequence}

Any tool that assumes Property~L---which includes \texttt{make},
\texttt{git}, \texttt{rsync}, \texttt{tar}, \texttt{sha256sum}, Time
Machine, and every backup utility in common use---will produce incorrect
results when operating on an iCloud Drive directory with evicted files.
The tool will believe it has processed all files in the directory; in
reality, it has processed only those files whose data extents happened
to be materialized at the moment of access.

\end{document}